\documentclass[runningheads]{llncs}
\usepackage{graphicx}
\usepackage{lipsum}
\usepackage{xcolor}
\usepackage{xspace}
\usepackage{soul}
\usepackage[hyphens]{url}
\usepackage{booktabs}
\usepackage{algorithm2e}
\usepackage{comment}
\usepackage{subcaption}
\usepackage{paralist}
\usepackage{color}
\usepackage{soul}
\usepackage{algorithmic}
\usepackage{graphicx}
\usepackage{textcomp}
\usepackage{acronym}
\usepackage{todonotes}
\usepackage{lipsum}
\usepackage[many]{tcolorbox}
\usepackage{cleveref}
\usepackage{listings}
\usepackage{xcolor}
\definecolor{delim}{RGB}{20,105,176}
\definecolor{numb}{RGB}{106, 109, 32}
\definecolor{string}{rgb}{0.64,0.08,0.08}

\lstdefinelanguage{json}{
    showspaces=false,
    showtabs=false,
    breaklines=true,
    postbreak=\raisebox{0ex}[0ex][0ex]{\ensuremath{\color{gray}\hookrightarrow\space}},
    breakatwhitespace=true,
    basicstyle=\ttfamily\small,
    upquote=true,
    morestring=[b]",
    stringstyle=\color{string},
    literate=
     *{0}{{{\color{numb}0}}}{1}
      {1}{{{\color{numb}1}}}{1}
      {2}{{{\color{numb}2}}}{1}
      {3}{{{\color{numb}3}}}{1}
      {4}{{{\color{numb}4}}}{1}
      {5}{{{\color{numb}5}}}{1}
      {6}{{{\color{numb}6}}}{1}
      {7}{{{\color{numb}7}}}{1}
      {8}{{{\color{numb}8}}}{1}
      {9}{{{\color{numb}9}}}{1}
      {\{}{{{\color{delim}{\{}}}}{1}
      {\}}{{{\color{delim}{\}}}}}{1}
      {[}{{{\color{delim}{[}}}}{1}
      {]}{{{\color{delim}{]}}}}{1},
}

\tcbset{
    sharp corners,
    colback = white,
    before skip = 0.2cm,    
    after skip = 0.5cm      
}      

\definecolor{main}{HTML}{5989cf}    
\definecolor{sub}{HTML}{cde4ff}     

\newtcolorbox{boxH}{
    colback = sub, 
    colframe = main, 
    boxrule = 0pt, 
    leftrule = 6pt 
}

\newcommand{\takeaway}[1]{\begin{boxH}\textbf{\textsc{Takeaway}}#1\end{boxH}}

    \title{The Impact of SBOM Generators on Vulnerability Assessment in Python: \\A Comparison and a Novel Approach}
\titlerunning{The Impact of SBOM Generators on Vulnerability Assessment in Python}

\author{%
Giacomo Benedetti\inst{1}  \and 
Serena Cofano\inst{1,2} \and
Alessandro Brighente\inst{3} \and \\
Mauro Conti\inst{3,4}
}%
\institute{
Department of Informatics, Bioengineering, Robotics and Systems Engineering, University of Genoa, Italy\\
\email{giacomo.benedetti@dibris.unige.it}\\ 
\and
IMT School for Advanced Studies Lucca, Italy\\
\email{serena.cofano@imtlucca.it}
\and 
Department of Mathematics, University of Padua, Italy\\
\email{\{alessandro.brighente,mauro.conti\}@unipd.it}\\
\and
Faculty of Electrical Engineering, Mathematics and Computer Science, Delft University of Technology, Netherlands
}

\newcommand{\filebased}{Metadata Based\xspace}
\newcommand{\envbased}{Environment Based\xspace}

\newcommand{\cdxgen}{cdxgen\xspace}
\newcommand{\syft}{Syft\xspace}
\newcommand{\ghsbom}{GH-sbom\xspace}
\newcommand{\trivy}{Trivy\xspace}
\newcommand{\ort}{ORT\xspace}

\newcommand{\sectool}{vulnerability scanner\xspace}

\newcommand{\sbomtool}{SBOM generation tool\xspace}
\newcommand{\pip}{PIP\xspace}
\newcommand{\tbd}{\textsc{\pip-sbom}\xspace}
\newcommand{\tbdmethod}{native\xspace}
\newcommand{\pipmodule}{\textsc{sbom}\xspace}

\newcommand{\grype}{Grype\xspace}

\newcommand{\vclause}[1]{\texttt{#1}}
\newcommand{\veq}{\vclause{==}}

\newcommand{\vle}{\vclause{\textless=}}
\newcommand{\vge}{\vclause{>=}}
\newcommand{\vlt}{\vclause{\textless}}
\newcommand{\vgt}{\vclause{>}}
\newcommand{\vneq}{\vclause{!=}}

\newacro{cisa}[CISA]{Cybersecurity and Infrastructure Security Agency}
\newacro{sbom}[SBOM]{Software Bill of Materials}
\newacro{ssc}[SSC]{Software Supply Chain}
\newacro{vex}[VEX]{Vulnerability Exploitability eXchange}
\newacro{purl}[purl]{package URL}
\newacro{sca}[SCA]{Software Composition Analysis}

\newcommand{\Rqone}{\textit{How much does the SBOM generation process impact the detection of vulnerabilities in the dependency network of an SSC?}}
\newcommand{\Rqoneone}{\textit{How does a specific vulnerability scanner perform when fed with an \ac{sbom} generated by a specific state-of-the-art \sbomtool{}s?}}
\newcommand{\Rqonetwo}{\textit{How much does an \ac{sbom} generation approach affect the performance of a \sectool?}}

\newcommand{\Rqtwo}{\textit{How can we improve the SBOM generation approach to achieve better performance on the security assessment of the dependency network in an SSC?}}

\begin{document}

\maketitle
\begin{abstract}
The Software Supply Chain (SSC) security is a critical concern for both users and developers.
Recent incidents, like the SolarWinds Orion compromise, proved the widespread impact resulting from the distribution of compromised software. 
The reliance on open-source components, which constitute a significant portion of modern software, further exacerbates this risk. 
To enhance SSC security, the Software Bill of Materials (SBOM) has been promoted as a tool to increase transparency and verifiability in software composition. 
However, despite its promise, SBOMs are not without limitations. 
Current SBOM generation tools often suffer from inaccuracies in identifying components and dependencies, leading to the creation of erroneous or incomplete representations of the SSC. 
Despite existing studies exposing these limitations, their impact on the vulnerability detection capabilities of security tools is still unknown. 

In this paper, we perform the first security analysis on the vulnerability detection capabilities of tools receiving SBOMs as input. We comprehensively evaluate SBOM generation tools by providing their outputs to vulnerability identification software. Based on our results, we identify the root causes of these tools' ineffectiveness and propose \tbd, a novel pip-inspired solution that addresses their shortcomings.
\tbd provides improved accuracy in component identification and dependency resolution. 
Compared to best-performing state-of-the-art tools, \tbd increases the average precision and recall by $60\%$, and reduces by ten times the number of false positives.

\keywords{Software Bill of Materials, Vulnerability Assessment, Dependency Network, Software Supply Chain Security}
\end{abstract}

\section{Introduction}
\label{sec:introduction}
The security of the \ac{ssc} is an increasing concern for users and developers as reported by both ENISA~\cite{enisa} and the UE Executive Order on Improving the Nation's Cybersecurity~\cite{biden}. 
Indeed, incidents such as the infection of SolarWind's Orion platform demonstrated how far-reached and impactful the distribution of compromised software is~\cite{peisert2021perspectives}.
The security of the \ac{ssc} depends on multiple factors, including, but not limited to, the use of open-source software as dependencies included in the developed application~\cite{ladisa2023sok}.
In a 2023 study of 1,703 commercial codebases across 17 industry sectors, Synopsys found that 96\% of them leverage open-source code, and 76\% of the total application code was open-source~\cite{OSSRA2023}.
Therefore, targeting software components, such as libraries, allows attackers to affect a wide range of software using a single entry point~\cite{ladisa2023sok,wermke2023always,guo2023empirical,merrill2023speranza}.

To improve the security posture of the \ac{ssc}, the Executive Order~\cite{biden} pushed the \ac{sbom} as a tool to increase the transparency and verifiability of the distributed software.
According to \ac{cisa}, an \ac{sbom} is ``\textit{a formal record containing the details and
supply chain relationships of various components used in building software}''~\cite{cisafaq},  hence providing developers and enterprises with transparency in the software composition.
An \ac{sbom} provides benefits for both software suppliers and consumers, as it helps identify and avoid known vulnerabilities, quantify and manage licenses, identify security and license compliance, and manage mitigation of vulnerabilities. 
\acp{sbom} are generated by automated tools and created according to different formats, with the most common being Software Identification (SWID) tagging, Software Package Data Exchange (SPDX), and CycloneDx~\cite{cisafaq}.
To fully leverage the information provided by \acp{sbom}, a plethora of tools have been developed to receive SBOM as input and provide security information~\cite{slscan-xj,grype,bomber-ru,kubeclarity-ue}.
To gather information on the security of components listed in the \ac{sbom}, these tools rely on open-source vulnerability databases, e.g., the NVD, to map components to vulnerabilities.
Software components can be associated with different information security depending on the database~\cite{dietrich2023security}. 
Sometimes, \acp{sbom} can also be complemented with a \ac{vex}, which provides additional information on possible specific vulnerabilities affecting software components of the \ac{sbom}.
Overall, the use of \acp{sbom} both increases the transparency of the distributed software and speeds up software adoption and testing. 
Indeed, retrieving known vulnerabilities of the listed software components via polling a public database is faster than running static or dynamic analysis application security testing.\\

\noindent{\textbf{Are SBOMs Improving Security?}
Although the premises are good, the \ac{sbom} is not what it is expected to be for security.
Most SBOM generation tools use specification files (e.g., setup.py, gemspec, package.json) to gather the dependency index. 
Other approaches are based on source or binary code parsing. 
Due to the lack of a standardized \ac{sbom} format and the limitations in the accuracy of existing \ac{sbom} generation tools~\cite{bi2024way}, it is possible to end up with different generated \acp{sbom} for the same software.
Indeed, the \ac{sbom} generation process depends on the tool's capability of correctly identifying components' names, versions, and dependencies. Wrongly identifying one or more of these elements impairs the representation capabilities of the resulting \ac{sbom}~\cite{bi2024way}.
Moreover, the claims made by \ac{sbom} generation tools on their support capabilities for different metadata file formats and their performance on real code bases are not consistent. 
Indeed, different and well-known \ac{sbom} generation tools are prone to parsing errors or inability to correctly gather all dependencies, resulting in an \ac{sbom} that represents a subset of the entire code base~\cite{deepbits,balliu2023challenges,rabbi2024sbom,Cofano2024-zp}. 
This represents a problem not only because the \ac{sbom} does not accurately represent the code, but also because missed key dependencies may lead to missed key security issues. Such dependency resolution problem is still an open issue for \ac{sbom} generation, and it is not clear how this impacts the security analysis capabilities provided via \acp{sbom}. 
Approaches such as code-centric call graph analysis and behavioral analysis may solve this problem, however, they are highly resource intensive~\cite{Hejderup2022,Plate2015,Ponta2018,Ponta2020}.\\

The limitations of the currently existing \ac{sbom} generation tools and the need for secure solutions to improve the security posture of the \ac{ssc} led us to the following research questions:

\begin{itemize}
    \item \textbf{RQ1}: \Rqone{}
    \begin{itemize}
        \item \textbf{RQ1.1}: \Rqoneone{}
        \item \textbf{RQ1.2}: \Rqonetwo{}
    \end{itemize}
    \item \textbf{RQ2}: \Rqtwo{}
\end{itemize}

\noindent{\textbf{Contributions.}}
In this paper, we evaluate for the first time how the representational capabilities of \ac{sbom} generation tools impact the identification of known vulnerabilities in the \ac{ssc}. 
Despite existing work evaluating \ac{sbom} generation tool according to their ability to correctly identify component's name, version, and dependencies for Java~\cite{balliu2023challenges}, JavaScript~\cite{rabbi2024sbom}, and Python~\cite{Cofano2024-zp}, no existing work evaluated their impact on the detection of security issues. 
As outlined in our research questions, we expect these issues to greatly impact on the identification of known vulnerabilities in the \ac{ssc}. 
To this aim, we selected five of the most relevant \ac{sbom} generation tools --- i.e., \cdxgen, \ghsbom, \ort, \syft, and \trivy--- for the evaluation.
Since many \ac{sbom} generation tools operate on the set of the most used programming languages (e.g., Python, JavaScript), we focus on the Python programming language, the most popular programming language in 2024 according to IEEE \cite{ieeerank}.
To solve the issues of \ac{sbom} generation tools and improve the security posture of the \ac{ssc} we propose \tbd, a novel \texttt{pip}-based solution. 
Our solution overcomes the most relevant issues of \ac{sbom} generation tool, i.e., it correctly identifies component names and versions and can correctly report all software dependencies.
We compare the performance of our solution with that of existing state-of-the-art tools and show that the \ac{sbom} generated with our tool drastically (64\% more precise than the best performing \sbomtool) increases the capabilities of identifying known vulnerabilities in the \ac{ssc}.

We summarize our contributions as follows.
\begin{itemize}
    \item We evaluate the capabilities of \ac{sbom} generation tools in helping increasing the \ac{ssc} security posture. By providing generated \acp{sbom} as input to a vulnerability scanner tool, we evaluate each \ac{sbom} generation tool in terms of the number of identified known vulnerabilities.

    \item We propose \tbd, an extension for \pip to generate an \ac{sbom} directly from the package manager, improving both usability and accuracy of the \ac{sbom} in the Python ecosystem.

\end{itemize}

\section{Background}
\label{sec:background}
This section provides the necessary background on the \ac{sbom} generation process (\Cref{sec:sbom-gen-proc}), the dependency management and resolution for Python (\Cref{sec:dep-man-python}), and the usage of \acp{sbom} by vulnerability scanners (\Cref{sec:vuln-scan-sbom}).

\subsection{SBOM Generation Process}
\label{sec:sbom-gen-proc}
An \ac{sbom} is generated using tools commonly known as \sbomtool{}s.
Differently from \ac{sca} tools, a \sbomtool does not analyse licenses and the security posture of components in the software under scrutiny.
However, it may be part of the \ac{sca}, providing inputs for further analysis of the identified assets.
\sbomtool{}s take as input the software's project folder and produce a list (i.e., the \ac{sbom}) of software components and their dependencies, along with their version and other information useful to trace the software composition.
As we focus on Python, in this paper, we showcase the \ac{sbom} generation approach for this language. 
However, the same process applies to all the other programming languages with just some differences related to their dependency resolution process.

Python has multiple package managers that a developer can choose to deal with dependencies and other project management operations.
Each package manager chooses how to deal with the project's filesystem to coordinate the dependency management.
Depending on the package manager, the project may result in very different filesystem structures.

\sbomtool{}s analyze the structure of a Python package or project and produce an \ac{sbom}.
How the \ac{sbom} is generated depends on the generation approach implemented by the \sbomtool{}.
Most tools rely on static metadata-based generation methods.
However, previous research~\cite{Yu2024,rabbi2024sbom} reported that \acp{sbom} have scarce accuracy when generated by tools currently implementing this approach.
Other tools, such as \cdxgen, aim to reproduce an installation environment where dependencies are collected according to metadata files.
NTIA established the minimum required elements for an \ac{sbom}~\cite{NTIAUnknown-vs}.
Tools such as sbom-scorecard~\cite{sbom-scorecard-pm} can quantify the level of compliance.
However, concerning vulnerability assessment, the required elements include only dependency identifiers --- i.e., name, version, and \ac{purl}.

\subsection{Dependencies Management in Python}
\label{sec:dep-man-python}
A Python project should contain either a \texttt{setup.py} or a \texttt{pypro\-ject.toml} file to be managed by a package manager.
Both these files contain the project's metadata and list the dependencies required to properly build and operate the project.
The project's build produces an artifact, a wheel or source distribution, containing the source code files and all the additional files required in the metadata file.

A distributable artifact is installed through \pip.
During installation, the dependencies listed in the metadata file are collected and installed on the user's system.
Since transitive dependencies --- i.e., dependencies of a dependency --- are not shipped together with the distributable artifact, \pip uses the \texttt{resolvelib} package implementing a specific algorithm for dependency resolution~\cite{pip-dep-solving}.

Dependencies are listed in the metadata files by name and version.
The dependency's version can be either pinned, non-pinned, or omitted --- i.e., not specified at all, in this case, the package manager collects the dependency to the latest stable and available version.
PyPI allows for multiple versioning schemas~\cite{python-versioning}, such as semantic versioning~\cite{Preston-WernerUnknown-qy}, calendar versioning~\cite{HashemiUnknown-by}, and their combinations.
The pinning vs. non-pinning choice is left to the developer by using the `\veq' operator vs. using range operators --- i.e., \vle, \vge, \vlt, \vgt, \vneq.

\subsection{Vulnerabilities Scanning with SBOM}
\label{sec:vuln-scan-sbom}
In this paper, we refer to \textit{vulnerability scanner} as a tool that analyzes a software artifact and provides a \textit{security report} listing potential vulnerabilities affecting the scanned product.

Analyzing a software's dependency network is not an easy task.
Modern software vastly relies on third-party software, resulting in the size of the dependency network rapidly increasing.
Vulnerability databases, such as NVD and OSV, contain entries for known software vulnerabilities, making the dependency network analysis easy and fast.
\acp{sbom} enable \sectool{}s to check the presence of known vulnerabilities without retrieving the dependency list.
Currently largely used \sectool{}s --- e.g., ShiftLeftScan~\cite{slscan-xj}, Grype~\cite{grype}, KubeClarity~\cite{kubeclarity-ue}, Bomber~\cite{bomber-ru} ---  use the \ac{sbom} as source for their analysis of dependencies.
They usually run a \sbomtool{} in the background and use the generated \ac{sbom} to resolve components names and retrieve their security information from public databases (e.g., NVD).

The use of \acp{sbom} makes the behavior of these \sectool{}s straightforward.
They parse the \ac{sbom} collecting dependency identifiers, such as \acp{purl}, and search for a match in vulnerability databases.

\begin{figure}[t]
\centering
\begin{lstlisting}[language=json]
{
  "matches": [{
    "vulnerability": { "id": "GHSA-9wx4-h78v-vm56", 
    "severity": "Medium", 
    "fix": { "version": "2.32.0" }},
    "relatedVulnerabilities": [{
    "id": "CVE-2024-35195", 
    "severity": "Medium" }],
    "matchDetails": { "type": "exact-direct-match", 
    "package": { "name": "requests", "version": "2.31.0" }, 
    "found": {
      "versionConstraint": "<2.32.0 (python)",
      "vulnerabilityID": "GHSA-9wx4-h78v-vm56"
     }
    "artifact": { 
    "name": "requests", 
    "version": "2.31.0", 
    "purl": "pkg:pypi/requests@2.31.0" }
  }]
}


\end{lstlisting}
\caption{Example of a condensed Grype scan report for a Python \ac{sbom}.}
\label{fig:security-report-example}
\end{figure}

In this paper, we use Grype to obtain security reports from \acp{sbom}.
A \grype's security report contains the following fields for each vulnerability:
\begin{itemize}
    \item \textit{Vulnerability}, information on the specific matched vulnerability (e.g. ID, severity, CVSS score, fix information, links for more information)
    \item \textit{RelatedVulnerabilities}, information pertaining to vulnerabilities found to be related to the main reported vulnerability, e.g., if the tool matches a vulnerability on GitHub Security Advisory, also the upstream CVE is reported. 
    \item \textit{MatchDetails}, the elements matching the vulnerability, such as the version constraints for which the vulnerability is matched.
    \item \textit{Artifact}, information about the location of the package within the directory, package type, licensing information, \acs{purl}, CPEs, etc.
\end{itemize}
\Cref{fig:security-report-example} shows an example of a Grype's security report for a Python \ac{sbom}.


\section{Experimental Setup}
\label{sec:research-design}

\begin{figure}
    \includegraphics[width=\linewidth]{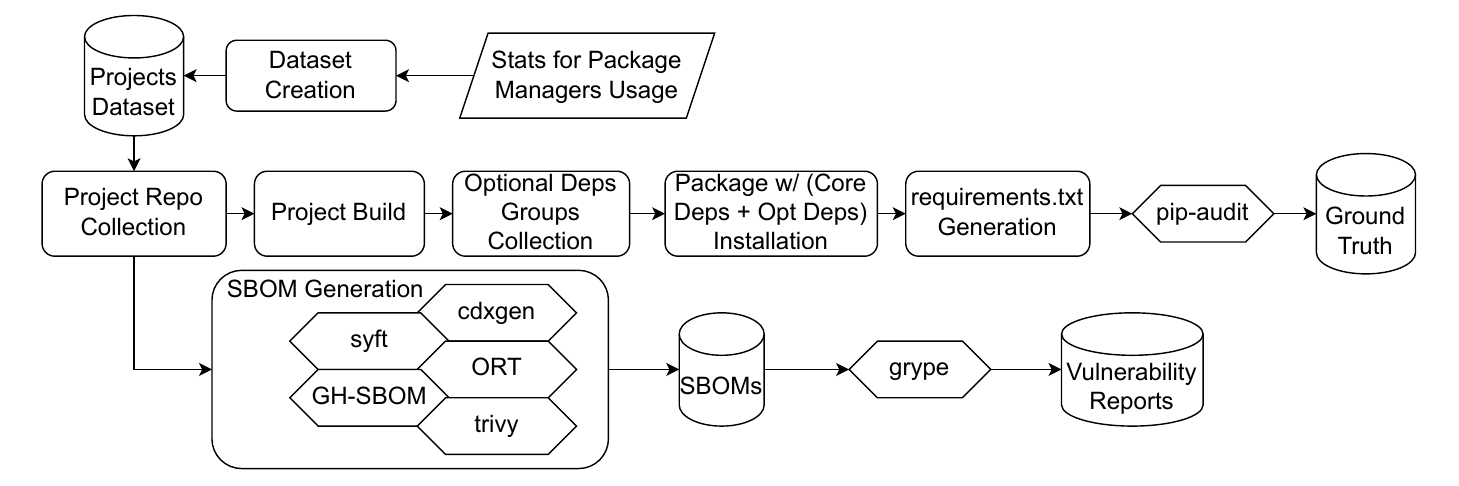}
    \caption{Experimental setup design. This approach provides us the necessary data to evaluate our research questions.}
    \label{fig:experimental-setup}
\end{figure}

In this section, we describe the setup for our evaluation methodology.
At first, \Cref{sec:projects-collection} provides the process we used to gather Python projects.
\Cref{sec:sbom-tools-selection} describes our selection of \sbomtool{}s based on their usage for security evaluation and their implemented generation method.
\Cref{sec:evaluation-process} reports the workflow applied to the collected projects to generate SBOMs and obtain their security reports.
\Cref{fig:experimental-setup} depicts the overall experimental setup.

\subsection{Projects Collection}
\label{sec:projects-collection}
Recalling from \Cref{sec:background}, Python allows the usage of multiple package managers, each implementing its way of dealing with dependencies.
Since \sbomtool{}s use specific parts of a Python project to generate the \ac{sbom}, we analyze tools' behavior in the context of different package managers.
To collect a representative sample of projects, we extract the distribution of the package managers used in 1,351 packages randomly selected from the whole population list on \url{ecosystem.ms}.
We discard packages that do not clearly state the package manager in their source code repository, obtaining the distribution of package managers shown in~\Cref{tab:package-manages-dist}.

\begin{table}[!h]
    \centering
    \caption{Package Managers and Their Usage}
    
    \begin{tabular}{p{4cm}  p{3cm} p{2cm}}
        \toprule
        \textbf{Package Manager} & \textbf{Packages} & \textbf{Percentage (\%)} \\
        \midrule
        poetry       &  38  &  6.44 \\
        pdm          &   9  &  1.53 \\
        hatch        &  85  & 14.41 \\
        pipenv       &   7  &  1.19 \\
        conda        &   0  &  0.00 \\
        setuptools   & 451  & 76.44 \\
        \bottomrule
    \end{tabular}
    
    \label{tab:package-manages-dist}
\end{table}

We collect a different sample of 1000 packages with the following process: 
\begin{inparaenum}[(1)]
    \item collect a random package from the entire package population hosted on PyPI;
    \item check the package manager used by the package;
    \item if we already reached the quota of packages using that specific package manager we discard the package, otherwise we add the package to the sample.
\end{inparaenum}
This process allows us to have a sample with the same proportion of package managers identified in the previous steps, and generalize our results to the entire package ecosystem with a 3.04\% margin of error at 95\% confidence level by standard sample size calculations.

\subsection{\sbomtool{}s selection}
\label{sec:sbom-tools-selection}
For the selection of \sbomtool{} we applied the following process:
\begin{inparaenum}[(1)]
    \item We manually scraped the list of tools on the CycloneDX tool center.\footnote{\url{https://cyclonedx.org/tool-center/}} We obtained a list of 169 open-source tools.
    \item We analyzed each tool in the list selecting those that generate \acp{sbom}, operate on Python, and have a command line interface. We reduce the list to 24 elements.
    \item We manually tested the 24 tools to prove their utility in this work. We excluded those that do not correctly execute, require external technologies (e.g., build-root), or were not maintained in the last year.
    Eventually, we obtained a list of 5 tools: \cdxgen, \ghsbom, \ort, \syft, and \trivy.
\end{inparaenum}

\Cref{tab:tools} lists the selected \sbomtool{}s, along with the implemented generation methodology and some example \sectool{}s making use of them.
The evaluation of the \acp{sbom} generated by these tools for a security assessment of the dependency network gives us the answer to RQ1.

\begin{table}[!h]
    \small
    \centering
    \caption{List of the selected \sbomtool{}s. Most of them are already officially used for dependency network security analysis. The selected tools can be also stratified based on the implemented generation method.}
    \begin{tabular}{p{3cm} p{3cm} p{4.5cm}}
    \toprule
     SBOM Gen. Tool & SBOM Gen. Met. & Example Sec. An. Tool  \\
    \midrule
        \cdxgen & \envbased  & Shiftleft Scan, Macaron~\cite{hassanshahi_macaron_2023}\\
       
        \syft & \filebased & Grype, KubeClarity\\
        
        \trivy & \filebased & KubeClarity \\
        
        \ort & \filebased & NA \\ 
       
        \ghsbom & Dep. Graph Based & NA \\
        \bottomrule
        
    \end{tabular}
    \label{tab:tools}
\end{table}

\subsection{Security Report Ground Truth}
A fundamental step to understand the effectiveness of \sbomtool{}s in generating an \ac{ssc} description that leads to the correct identification of vulnerabilities, is knowing the vulnerabilities that affect a specific project.
To obtain this ground truth, we follow a multistep approach.
For each package in our collected sample we automatically:
\begin{inparaenum}[(1)]
    \item retrieve the package's project from its code repository;
    \item parse metadata files to obtain optional dependency groups;
    \item install the package in a virtual environment along with both required and optional dependencies;
    \item generate the requirements.txt file using the \texttt{pip freeze} command to filter out packages installed in the virtual environment by default;
    \item pass the requirements.txt to pip-audit;
    \item collect the security report.
\end{inparaenum}
Thanks to this manual approach, we build the list of vulnerabilities associated with each project. A perfect \sbomtool{} will create a project representation the leads to the correct identification of this precise set of vulnerabilities.

\subsection{SBOMs and Security Reports Generation}
To generate relevant \acp{sbom}, we feed our selected \sbomtool{}s with the packages collected in our dataset. 
We parse \acp{sbom} with \texttt{jq}~\cite{jq-tn} (a JSON parser) to verify they have the expected format.
That is we verify they do not contain the metadata pointing out the correct analysis of the project by the \sbomtool{}.

We selected \grype~\cite{grype} as the tool for the generation of security reports.
\grype is a vastly used tool for the security analysis of projects~\cite{chainguard-grype-mu,cisa-grype-kq,Wallen2022-zn,chainguard-grype2-ap}.
It covers multiple languages --- e.g., Python, Go, Rust --- and artifacts --- e.g., Docker images, filesystems, \acp{sbom}.
As we do not need specific security analysis tool for this work, the tool just needs to parse the \ac{sbom} and query vulnerability databases for known vulnerabilities.
\grype queries multiple databases, cross-checking vulnerabilities, and hence represents a perfect choice for our purposes.

A \grype's security report contains matches for found vulnerabilities, allowing us to compare the set of found vulnerabilities with the set of vulnerabilities reported in the ground truth.

\section{\tbd: Our Proposed SBOM Generator}
\label{sec:poc}

This Section describes \tbd, our PIP-based approach for \tbdmethod generation method. Figure \ref{fig:pip-sbom} depicts its main steps.
\begin{figure}
    \centering
    \includegraphics[width=\linewidth]{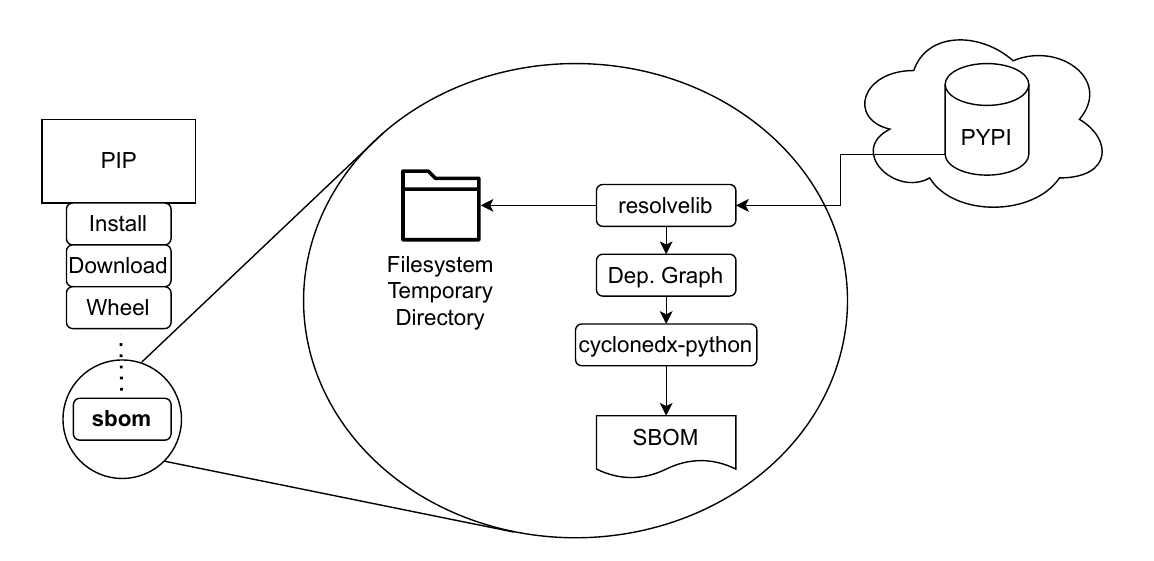}
    \caption{Design of \tbd. We extend the implementation of \pip to include \ac{sbom} generation in the build phase.}
    \label{fig:pip-sbom}
\end{figure}

\tbd is designed as part of PIP, the PyPI official package manager.
We extend this specific package manager because:
\begin{inparaenum}[(1)]
    \item it supports multiple front-ends and back-ends, i.e., it can build other package manager projects, such as poetry.
    \item Both skilled and novel developers commonly use it.
\end{inparaenum}

\pip is internally based on modules representing the possible CLI commands for a user.
We added a module --- i.e., \pipmodule --- containing the logic needed to generate the SBOM\footnote{\tbd is the extended version of \pip, while \pipmodule is the specific module extending \pip. Hereafter, we refer to both of them as \tbd, for simplicity of language.}.
This allows developers to generate an \ac{sbom} for a Python project with the command: \texttt{pip sbom <project-path>}.

\tbd includes an online process and an offline process.
The online process interacts with the PyPI registry obtaining the dependency network.
The offline process builds the dependency network graph and generates the SBOM document from a Python project.

\subsubsection{Dependency Network Solving}
\pip uses the \texttt{resolve-lib} package dealing with dependencies and version constraints.
This package optimizes the solving algorithm with the optimal navigation path of the dependency tree~\cite{resolvelib-ti}.
We build upon the logic already implemented in \pip for this package to mimic the same solving algorithm used during dependencies retrieval.

This process has a similar behavior to the \texttt{download} command.
The project's dependencies are collected from PyPI and stored inside a directory specified by the user.
In our implementation, the dependencies are stored inside a temporary directory and removed at the end of the process.
This process automatically solves version constraints similarly to the process happening during project build and installation, providing a reliable representation of the dependency network at installation time.
When a constraint cannot be solved because of version incompatibility, it is discarded, as would happen during the project installation.

The generation of the dependency graph is coupled with this process.
We decided to store collected dependencies as a graph to allow deeper investigation of dependency relationships when required.

\subsubsection{Dependency Network Graph}
The dependency network is defined as a direct unweighted graph $G=(V, E)$. 
Each element $n\in V$ is either a direct or transitive dependency of the input project $r\in V$.
An edge $(u, v)\in E$ represents a dependency relationship between the nodes $u$ and $v$.
This kind of dependency is a transitive binary relation.
That is, from $u \rightarrow v$ and $v \rightarrow w$ it follows $u \rightarrow w$.
In this case, we say that $u$ is a transitive dependency of $w$.

\tbd outputs the generated graph as a \texttt{dot} file when required through the \texttt{-g <file-name>} option.
The \tbd internally uses the dependency graph to generate the \ac{sbom}.

\subsubsection{SBOM Generation}
Once the graph is complete, \tbd module navigates the graph and creates an entry in the \texttt{components} field of the \ac{sbom} for each node.
An entry contains: the bom-ref, dependency name, version and purl.
We include only this information because they are necessary to the \sectool.
In general, the \ac{sbom} can be enriched to comply with the minimum required elements stated by NTIA~\cite{NTIAUnknown-vs}.

Edges of the dependency network are used to fill the \emph{dependencies} field of the \ac{sbom}, which contains the relationship between components.
When the graph exploration ends, \tbd produces a CycloneDx-compliant \ac{sbom}.

\section{Evaluation Methodology}
\label{sec:evaluation-process}
In this section, we present our methodology to answer our research questions.
The evaluation process is divided into two parts.
The first looks at the accuracy the \sectool reaches with \acp{sbom} generated by selected \sbomtool{}s.
The second looks at data specific to \tbd to evaluate how an \ac{sbom} generated with a different method affects the \sectool performance.

\subsubsection{RQ1: \ac{sbom} Generation Impact on Vulnerability Scanning}
This research question aims to understand to what extent the \ac{sbom} impacts the security analysis tool output.
The SBOM should accurately represent the SSC.
Thus, accurately representing the SSC is an enabling property for security analysis of the software dependency network.

\paragraph{RQ1.1: \Rqoneone{}}
To answer this question, we compare the vulnerabilities identified starting from a \acp{sbom} against the ground truth obtained through pip-audit.
We use the Jaccard similarity index to compute the degree of overlap and commonality among the two vulnerability sets.
For each tool and each SBOM $\mathcal{S}$, the process involves: 
\begin{inparaenum}[(1)]
    \item collecting the security reports generated from $\mathcal{S}$ and extract the matched vulnerabilities (\texttt{ToolVulns});
    \item getting the vulnerabilities stored in the ground truth for the project associated with $\mathcal{S}$ (\texttt{GrTrVulns});
    \item compute the Jaccard similarity index between \texttt{ToolVulns} and \texttt{GrTrVulns}, according to \Cref{eq:jaccard}.
\end{inparaenum}

\begin{equation}
J(ToolVulns, GrTrVulns) = \frac{|ToolVulns \cap GrTrVulns|}{|ToolVulns\cup GrTrVulns|} 
\label{eq:jaccard}
\end{equation}

\paragraph{RQ1.2: \Rqonetwo{}}
While Jaccard similarity represents a useful metric to assess the extent a tool suits security purposes, it cuts off details on the reasons behind the tool's performance.
To obtain these missing details and answer \textit{RQ1.2}, we compute the false positives, false negatives, precision, and recall.

Specifically, \textit{precision} assesses the fraction of correctly identified vulnerabilities, according to the ground truth, among all identified vulnerabilities (\Cref{eq:precision}).
\begin{equation}
Precision = \frac{TP}{TP+FP}
\label{eq:precision}
\end{equation}

\textit{Recall} measures the fraction of correctly identified vulnerabilities to all actual vulnerabilities in the ground truth (\Cref{eq:recall}).

\begin{equation}
Recall = \frac{TP}{TP+FN}
\label{eq:recall}
\end{equation}

These metrics provide a holistic overview of the performance of each tool's generation method: precision addresses the trustworthiness of identified vulnerabilities, and recall looks at the generation methods' efficacy in allowing security assessment to pinpoint pertinent vulnerabilities.

\subsubsection{RQ2: A Better \ac{sbom} Generation Approach}
As we later show, the main issue with currently existing SBOM generation approaches and the resulting poor performance of vulnerability identification approaches is that SBOM generation tools are not able to correctly identify components and their dependencies.
To answer \textit{RQ2} and improve the security posture of the SSC,
we extend PIP to investigate to what extent applying the same logic used in the retrieval of dependencies during a project installation for \ac{sbom} generation is beneficial.
The extension leverages the already present \texttt{download} module in PIP.
This module helps to download the package's archives without installing them.
Modifying this module by adding functional elements for \ac{sbom} generation provides us the means for creating a novel and better approach.
We provide the details of our implementation in \Cref{sec:poc}.

To evaluate the improvement in the security analysis by using a \ac{sbom} generated with our approach, we apply the same metrics used to answer RQ1.


\section{Evaluation Results}
\label{sec:results}
In this Section, we present the results of our analysis organized by research question.
In particular, Section \ref{sec:rq1eval} presents the results of \textit{RQ1}, while Section \ref{sec:rq2eval} presents the results of \textit{RQ2}.

\subsection{RQ1: \ac{sbom} Impact on Vulnerability Scanning}\label{sec:rq1eval}
Our findings show that the \acp{sbom} generation approach deeply impacts the performance of vulnerability scans.
All the analyzed \sbomtool{}s cannot lead to the correct identification of all vulnerabilities in more than 20\% of cases, with the only exception of \cdxgen achieving correct identification in almost 40\% of cases.

\subsubsection{RQ1.1: Impact of tools on vulnerability scanner performance}
The security reports generated from state-of-the-art \sbomtool{}s reveal some shortcomings, particularly regarding thoroughness and accuracy in identifying vulnerabilities.
The distribution of Jaccard similarity indexes in \Cref{fig:jaccard-distribution-pip-ext} provides a perspective of the true positive vulnerabilities found with \acp{sbom} generated by different tools.
From the results, it is clear that \sbomtool{}s badly impact on the vulnerability scans operated by \sectool{}s.

\begin{figure}[t]
    \centering
    \includegraphics[width=\linewidth]{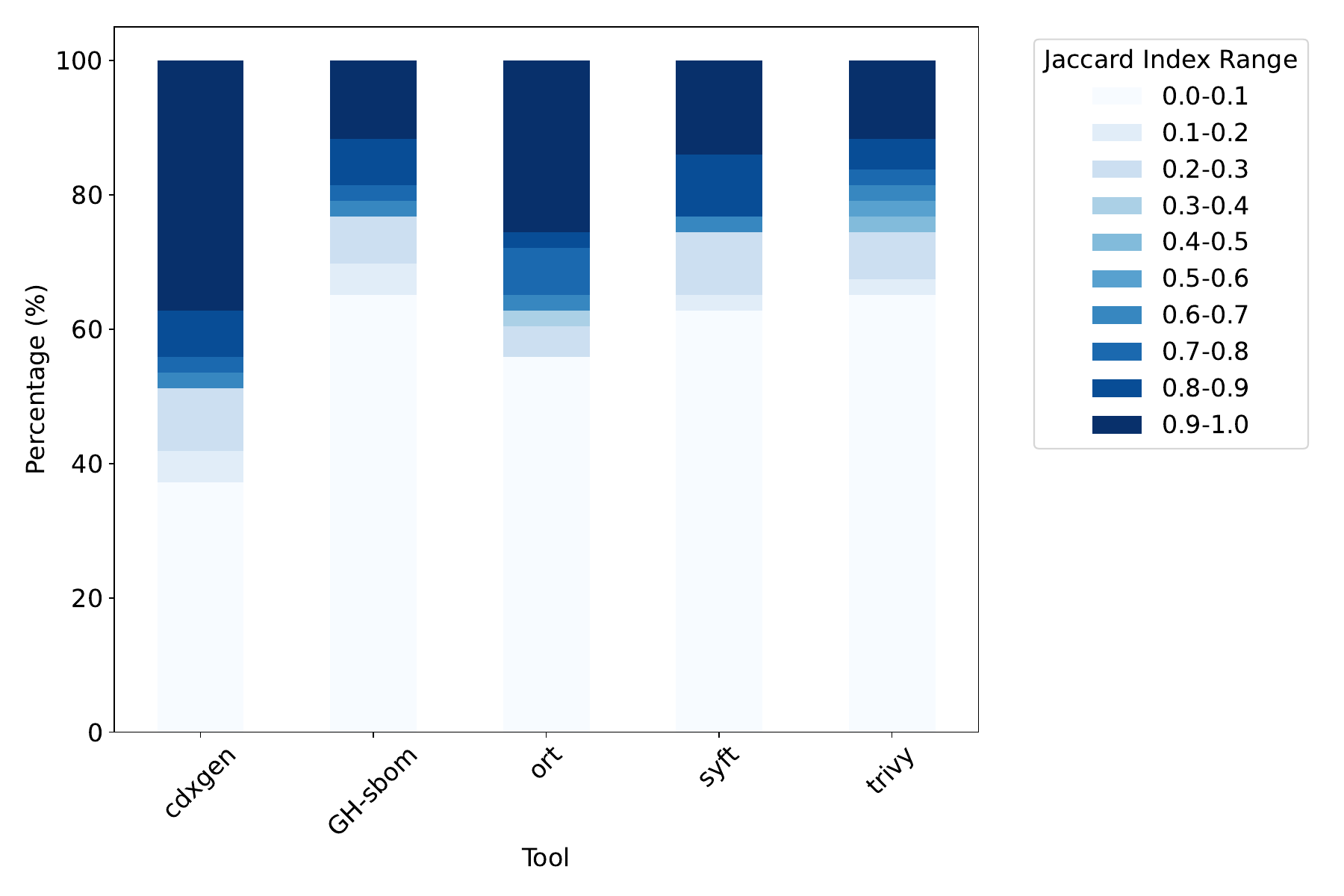}
    \caption{Jaccard Similarity Distributions. Each bar represents the percentage of SBOMs that lead to identification with a certain Jaccard index range. We include the vulnerability assessment obtained with \acp{sbom} generated with \pip for comparison purposes.}
    \label{fig:jaccard-distribution-pip-ext}
\end{figure}

Among the analyzed \sbomtool{}s, the one providing the best vulnerability scan results is \cdxgen.
This is motivated by the fact that \cdxgen 
\begin{inparaenum}[(1)]
    \item installs dependencies in a virtual environment, and 
    \item supports most of the package managers.
\end{inparaenum} 
These two properties result in being critical for a \sbomtool for a proper representation of the \ac{ssc}.
Hence, all tools lacking at least one of these two capabilities result in worse vulnerability scans.

\ort uses an approach similar to dependencies installation in a simulated environment.
That is, it uses an external Python module to query the PyPI registry and obtain information on dependencies.
However, its support for package managers is limited to Poetry and Pipenv, or projects including \texttt{requirements.txt} files.

\ac{sbom} generators solely relying on static metadata --- i.e., \syft and \trivy --- display worse performances for vulnerability detection.
This behavior is caused by the fact that they do not install dependencies, while they have good support for different package managers.
\ghsbom results are highly influenced by the settings applied by repository owners.
Thus, \ghsbom works \textit{only} when the dependency graph feature is enabled on the repository.
However, its main issue is that it targets the last commit on the main branch to generate the \ac{sbom}.
Hence, \ghsbom cannot acquire an \ac{sbom} for a specific commit or tag~\cite{ghsbom-issue-commit}, causing \sectool{}s to analyze \acp{sbom} belonging to code different from the targeted one.

\takeaway{
Current state-of-the-art \sbomtool{}s are not suitable to generate \acp{sbom} for the proper vulnerability assessment of Python projects.
}

\subsubsection{RQ1.2: Causes of tool impact on vulnerability scanner}
\begin{figure}[t]
    \centering
    \includegraphics[width=\linewidth]{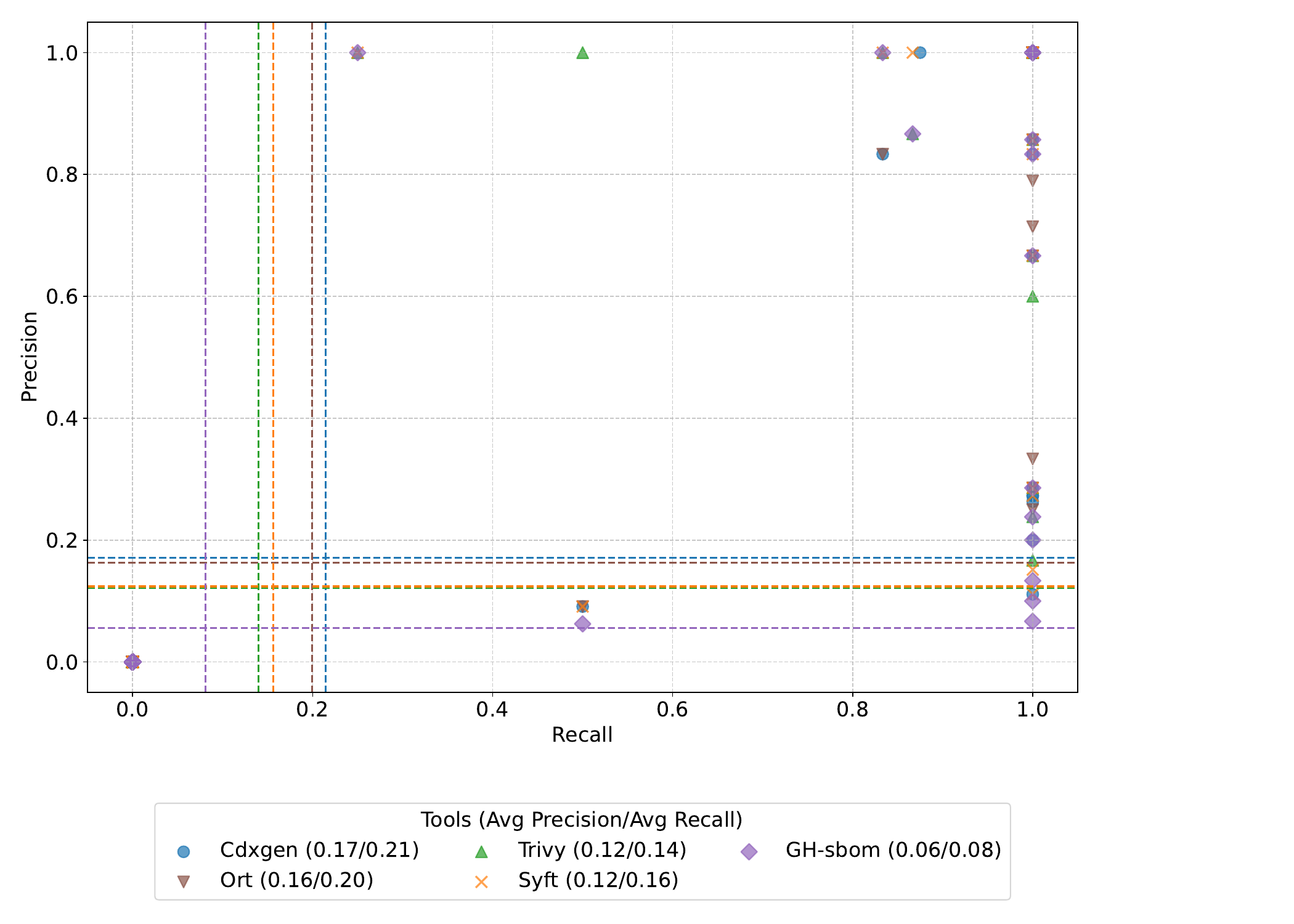}
    \caption{Precision and Recall for the vulnerability scans conducted through \acp{sbom} generated by each of the selected \sbomtool{}s.}
    \label{fig:precision-recall}
\end{figure}

While the Jaccard similarity represents how much the \ac{sbom} generation approach impacts the vulnerability scan results, precision and recall measurements give us information on the factors influencing the security evaluation.
Referring back to \Cref{sec:evaluation-process}, recall is the fraction of correctly identified vulnerabilities to all actual
vulnerabilities in the ground truth.
Precision is the fraction of correctly identified vulnerabilities, according to the ground truth, among all identified vulnerabilities.
\Cref{fig:precision-recall} shows the precision and recall values for the analyzed \ac{sbom} generation tools.
All the tools show low precision and recall average values, with \cdxgen leading the group with 0.17 and 0.21 average precision and recall, respectively.

\begin{figure}[t]
    \centering
    \includegraphics[width=\linewidth]{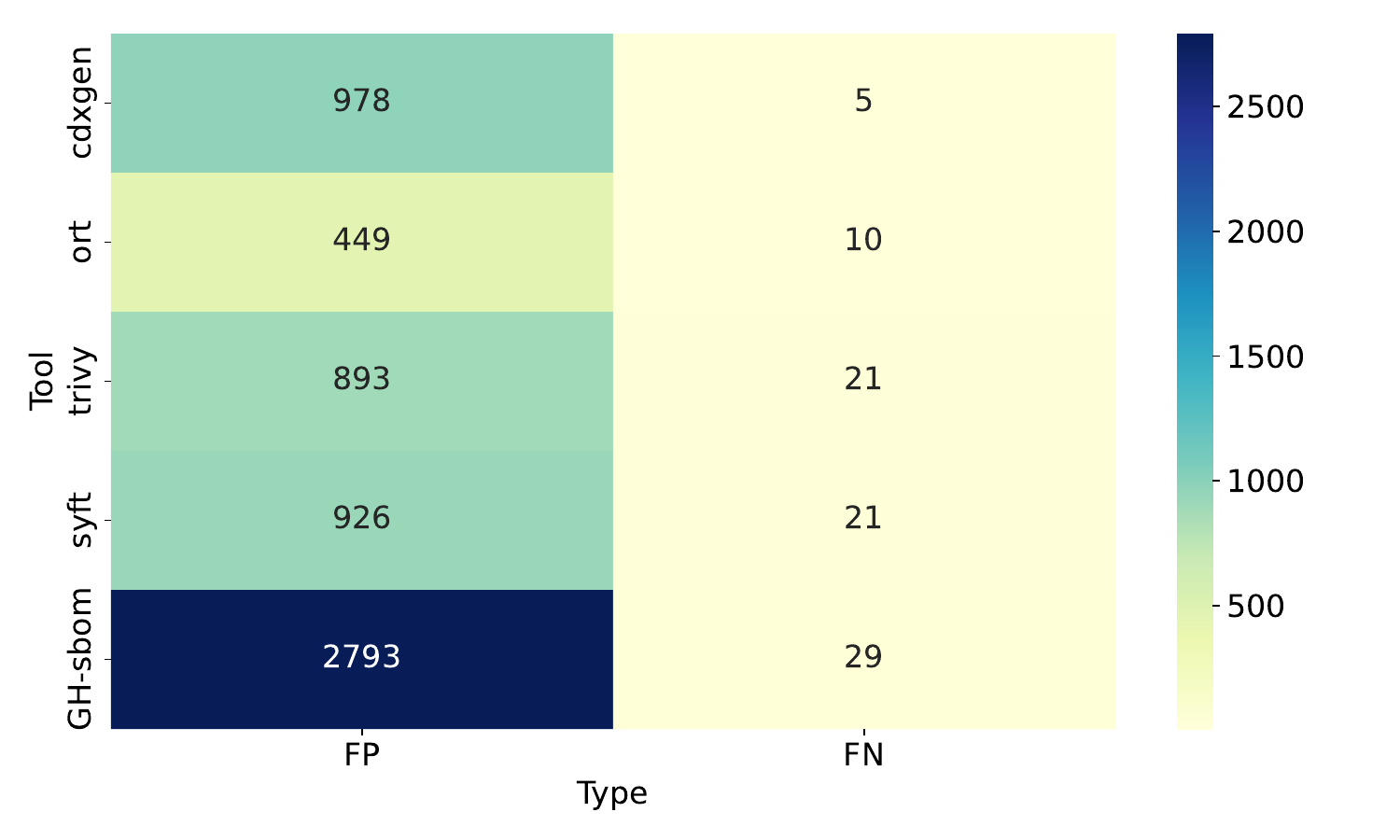}
    \caption{False Positives and False Negatives by Tool}
    \label{fig:heatmap-fp-tp}
\end{figure}

As shown in~\Cref{fig:heatmap-fp-tp}, the analyzed tools have a very high number of false positives, while false negatives are present in a more manageable magnitude.
For example, 99.5\% (978 FP / 5 FN) of the misclassified vulnerabilities are false positives for \cdxgen and 97.8\% (926 FP / 21 FN) for \syft.
While an overestimation is generally considered better than losing vulnerabilities~\cite{Owasp-false-positives-cf,guardrails-false-positives-zy,contrast-false-positives-uc}, these numbers are out-of-scale, causing burdening during vulnerability investigation.

By randomly sampling projects with at least a false positive we identified the following reasons:
\begin{enumerate}
    \item The \ac{sbom} list dependencies that are not collected during installation,
    \item the vulnerability is reported with a vulnerability identifier different from the one in the ground truth
\end{enumerate}

The first reason is caused by the presence of metadata files inside of projects listing dependencies that are \emph{not} actually collected during installation.
By analy\-sing the dependencies contained in the \acp{sbom} we confirm that on average 75\% of the dependencies listed in the \acp{sbom} are not actually installed. 
When a package is built only the files specified inside of the metadata file are included in the package (see \Cref{sec:dep-man-python}).
Those are the dependencies that the package manager collects during the installation.
This misalignment between files listing dependencies, and files used for installing dependencies, causes \sbomtool{}s to generate a huge amount of false positives during vulnerability scanning.

The second reason is linked to a limitation of our methodology (see \Cref{sec:limitations}) and affects a limited number of vulnerabilities making it negligible.
However, it highlights the problem of using vulnerability databases as the source for vulnerability scanning.
These databases may be partially out-of-date, or experiencing problems in the collection of security advisories.
Recently NVD had trouble collecting CVEs for a long time, causing many issues for tools relying on such database~\cite{nvd-problem}.

\takeaway{
\begin{inparaenum}[(1)]
    \item \sbomtool{}s do not properly support multiple package managers.
    \item \sbomtool{}s do not consider how dependencies are actually collected by Python's package managers.
\end{inparaenum}
}

\subsection{RQ2: Trying a Different \ac{sbom} Generation Method}\label{sec:rq2eval}

\begin{table}[t]
\def\arraystretch{1.5}
    \caption{Comparison of average values for Jaccard similarity, Precision, and Recall for \tbd against state-of-the-art tools.}
    \centering
    \begin{tabular}{lrrrrr|r}
    \toprule
      & \cdxgen & \ort & \syft & \trivy & \ghsbom & \textbf{\tbd} \\
     \midrule
     Jaccard Similarity   & 49.77\% & 36.50\% & 26.33\% & 23.63\% & 23.98\%  & \textbf{78.39\%} \\
        Avg Precision & 17.08\% & 16.31\% & 12.39\% & 12.17\% & 5.57\% & \textbf{80.95\%} \\
        Avg Recall  & 21.42\% & 19.93\% & 15.61\% & 14.01\% & 8.10\% & \textbf{80.26\%} \\
        F. Poss. / F. Negs. & 978/5 & 449/10 & 926/21 & 893/21 & 2793/29 & \textbf{47/3}\\
        \bottomrule
    \end{tabular}
    \label{tab:pip-sbom-results}
\end{table}

Implementing \tbd, a \pip extension using the dependency resolution algorithm native to the package manager, we drastically improve the vulnerability assessment of the dependency network.
As \Cref{tab:pip-sbom-results} reports, almost 80\% of the vulnerability reports match the ground truth.
\emph{No one of the analyzed pre-existing tools is able to provide the same performance.}

\tbd achieves a 64\% increase in precision and a 59\% improvement in recall for vulnerability scans, outperforming the best-performing existing tool.

Having a limited delta between average precision (80.95\%) and recall (80.26\%), our proposed tool allows a \sectool to effectively identify vulnerabilities in the dependency network requiring a limited manual effort to discard false positive vulnerabilities. 
Thus, it has only 47 false positive vulnerabilities.
While the false negative vulnerabilities have the same magnitude as the other \sbomtool{}s, false positives are way lower than other tools.
By providing developers with a manageable number of vulnerabilities to check, we want to push towards the adoption of \acp{sbom} as a useful means for security.

Concerning projects' security assessment differing from the ground truth:
We manually reviewed these differences, and all of them are due to issues with vulnerability identifiers (See \Cref{sec:limitations} for details).

\takeaway{
\pip can be extended by re-using most of its code to generate an \ac{sbom} that drastically improves the vulnerability assessment results.
}

\section{Discussion}
\label{sec:Discussion}

\ac{sbom} generation is a hard problem.
Since software is usually constructed on third-party components, having a complete and correct \ac{sbom} represents a great improvement for security, and many other aspects of software usage --- e.g., licensing.
Using \ac{sbom} as input for \sectool{}s speeds up the analysis of the software's dependency network, also providing clear information on the vulnerable dependencies and their transitive dependencies.
However, state-of-the-art tools do not provide \acp{sbom} that can be efficiently used for vulnerability assessment in the Python ecosystem.

We identified two main causes for this problem:
\begin{enumerate}
    \item \sbomtool{}s do not provide support for the high number of package managers used for Python projects.
    \item They do not correctly build the dependency network.
\end{enumerate}

However, it is possible to greatly improve the current situation, without much of an effort.
Since dependencies are collected by package managers, using them for \ac{sbom} generation represents an efficient approach.
We tested this \textit{\tbdmethod} generation method by implementing it in \pip.

\subsection{Implications}
Our results can be interpreted in two ways:
\begin{itemize}
    \item Developers currently relying on tools like shiftleft scan or \grype are missing out most of the actual vulnerabilities in the dependencies of the analyzed software.
    \item The problems affecting the \sbomtool{}s can be easily solved by adding just some changes to package managers. Once the \ac{sbom} is correctly generated, it greatly benefits the vulnerability assessment.
\end{itemize}

These implications can be easily transferred to other languages and ecosystems.
Most sofware ecosystems do not support provenance and \ac{sbom} generation.
According to OpenSSF only npm ships the generation of SBOM for the hosted packages, while only npm and homebrew provide provenance information~\cite{package-repos-security}.

On the other hand, having a proper \ac{sbom} enables great result in the vulnerability assessment.
With just a few changes to the package manager, the \ac{sbom} can be shipped with a project.

\subsection{Recommendations}
Our recommendations are addressed to communities of software ecosystems.
There is a need to include \acp{sbom} as part of projects by default.
Thanks to our proposal, we showed that this is a possible and easily achievable goal.
The following recommendations may help those communities:
\begin{itemize}
    \item Using centralized \ac{sbom} may provide a common knowledge base, easily accessible, and distributed for all developers. \ghsbom provides such a feature, however, it is not always supported and it is limited to providing \ac{sbom} for the latest commit on the main branch.
    Fostering discussion on having such a powerful tool may relieve some of the efforts on the single communities.
    \item Push for a standard set of files to list the dependencies installed in a project. Most ecosystems already use this approach, for example, npm (\texttt{package.json}), and Cargo (\texttt{Cargo.toml}).
    Multiple data sources may be confusing and can introduce unexpected dependencies and associated vulnerabilities.
    \item Work for a \sbomtool{} implemented inside of the ecosystem package manager(s). As we showed, \ac{sbom} generation can largely be improved by using a \tbdmethod generation method. \acp{sbom} can benefit from ecosystem-specific information that may improve \ac{ssc} transparency for that specific ecosystem. 
\end{itemize}

\section{Threats to Validity}
\label{sec:limitations}
\paragraph{Projects Collection.}
Our sample can be generalized to the whole package population on PyPI, with the margin of error stated in \Cref{sec:projects-collection}.
While it is relevant because of generalization, it may miss specific corner cases providing insightful knowledge.
We internally cross-validated the sample with other random samples taken from PyPI without filtering on package managers.
The cross-validation confirmed the accuracy of our results with an error of $\pm2\%$.

\paragraph{\sbomtool{}s Selection.}
The selection is manually conducted by filtering \sbomtool{}s based on the criteria discussed in \Cref{sec:sbom-tools-selection}.
The list of tools has been reviewed by more than one author, and we agreed on the five selected tools.
However, a degree of subjectivity would be present in the selection, leading to the exclusion of other potentially effective tools.
We tested the tools that were eligible for our study and excluded the ones that were not functional for our research goals.
The manual testing may have caused the exclusion of potentially effective tools.

\paragraph{Ground Truth.}
The reliance on pip-audit for establishing the ground truth introduces potential limitations, as this tool may not detect all relevant vulnerabilities, which could impact the baseline used for comparing other tools.
pip-audit is largely used to detect vulnerabilities in the dependency installed inside of an environment. 
However, it is subject to vulnerability databases as well as \sectool{}s.
Since our measurements with \grype were conducted at the same time as the ground truth generation, we argue that any potential gap was avoided.

\paragraph{Evaluation Methodology.}
We experienced false positives and negatives while evaluating vulnerability scan results.
Some are due to a missing vulnerability identifier inside the vulnerability scan result.
These errors can be easily fixed by establishing a unique vulnerability database mapping vulnerabilities to their identifiers on the various databases.
We manually fixed the issue in our dataset since such an event is rare.

\paragraph{Evaluation Scope.}
The methodology and results are tailored to Python, while we consider these results extensible to other languages, we cannot state their transferability.
Our study wants to raise concerns about the reliability of current \sbomtool{}s for security analysis of dependency networks, and to push on the development of native \ac{sbom} generation method by package managers.

\section{Related Works}
\label{sec:related-works}
 
Research produced a large amount of literature on \ac{sbom} in the last period.
Related to this work, it can be divided into two categories: research on (1) technical challenges and (2) adoption.

\paragraph{Technical Challenges}
Yu et al.~\cite{Yu2024} conduct a differential analysis examining the correctness of \acp{sbom} generated by four \sbomtool{}s.
The analysis is conducted over seven program languages.
They highlight how the \sbomtool{}s have difficulties to correctly finding dependencies.

Torres\--Arias et al.~\cite{Torres-Arias2023-hx} conduct a study on the fulfilment of minimum required elements issued by NTIA~\cite{NTIAUnknown-vs} for \acp{sbom} using the SPDX standard.  
Similarly, Halbritter and Merli~\cite{Halbritter2024-mh} do the same for CycloneDX \acp{sbom}.

Balliu et al.~\cite{balliu2023challenges} focus on the Java ecosystem with the analysis of the \ac{sbom} generated for a known Java application.
They provide an overview of the challenges that \sbomtool{}s face to generate an \ac{sbom} on Java projects.
Similarly, Rabbi et al.~\cite{rabbi2024sbom} focus on the npm ecosystem.
Cofano et al.~\cite{Cofano2024-zp} conduct a study on the relationship between the Python ecosystem and four \sbomtool{}s.
They identify challenges posed by the ecosystem and an excess of approximation by the \sbomtool{}s.
All these works do not look at the impact of the generated \ac{sbom}. Differently, we focused on the usage of the \ac{sbom} to understand to what extent this technology can be effectively adopted.

\paragraph{\ac{sbom} adoption}
While the \ac{sbom} represents a resource for \ac{ssc} transparency, enabling both functional and security testing, its adoption is delayed.
The efficacy of \ac{sbom} for security has been recently reported by Sharma et al.~\cite{sharma2024sbomexecounteringdynamiccode}.
They propose a technology using \ac{sbom} to mitigate vulnerabilities affecting Java applications.

Enck et al.~\cite{Enck2022-ox} report that the use of \ac{sbom} for security is debated among practitioners.
More than a year later the landscape is not brighter.
Zahan et al.~\cite{Zahan2024-ez} report that attended of the S3C2~\cite{s3c2} Industry Summit were sceptical about the adoption of \ac{sbom}.
The problems come from including \ac{sbom} generation in the CI/CD pipeline, however, it has been suggested \emph{to embed \ac{sbom} generation within the build template as part of a standardized build pipeline and making SBOM generation a mandatory task when setting up CI/CD templates}.
We show that this approach can be easily adopted without breaking the build process, at least concerning the Python ecosystem.

\section{Conclusion and Future Work}
\label{sec:Conclusion}
This study shows how \ac{sbom} generation heavily affects vulnerability scans of software.
The current state-of-the-art \sbomtool{}s cannot provide \acp{sbom} accurate enough for \sectool{} to identify more than 20\% of the vulnerabilities actually present in the software.
This problem is due to the lack of support for Python and the techniques adopted to solve versions of Python project dependencies.
We proposed \tbd, a proof of concept implementation extending \pip to generate an \ac{sbom} directly from the Python package manager.
The performances provided by our PoC for vulnerability assessment suggest that using native resources of the ecosystem --- e.g., the package manager --- to generate the \ac{sbom} may largely improve the overall security posture of software. 

Some future works in this direction can support the generation of \ac{sbom} in different software ecosystems.
Moreover, research in the use of \ac{sbom} for security should provide further elements to push software communities and developers to adopt this technology.

\section{Data Availability}
The list of projects, ground truth, generated \acp{sbom}, security reports, scripts to replicate our results, and \tbd will be released upon publication.

\bibliographystyle{splncs04}
\bibliography{main.bib} 

\begin{thebibliography}{10}
\bibitem{kubeclarity-ue}
kubeclarity: {KubeClarity} is a tool for detection and management of software bill of materials ({SBOM}) and vulnerabilities of container images and filesystems
\bibitem{slscan-xj}
sast-scan: Scan is a free \& open source {DevSecOps} tool for performing static analysis based security testing of your applications and its dependencies. {CI} and git friendly
\bibitem{sbom-scorecard-pm}
sbom-scorecard: Generate a score for your sbom to understand if it will actually be useful
\end{thebibliography}

\end{document}